# Lexical Anthropomorphization Influences on Moral Judgments of AI Bad Behavior

Jaime Banks
School of Information Studies
Syracuse University
Syracuse, NY, USA
banks@syr.edu

Nicholas David Bowman
Newhouse School
Syracuse University
Syracuse, NY, USA
nbowman@syr.edu

Roman Saladino
School of Information Studies
Syracuse University
Syracuse, NY, USA
rasaladi@syr.edu

## ABSTRACT

Anthropomorphic language describing artificial intelligence (AI) is widespread in media, policy, and everyday discourse; so too are discussions of AI bad behavior, from hallucinations to inappropriate comments. How does humanizing language about AI shape moral judgments when AI behaves badly? Across four experiments (total $N$ = 1,020), we tested whether lexical anthropomorphism (LA) primes shape judgments of AI moral character, behavior morality, and behavioral responsibility. Studies 1–3 tested interactions between anthropomorphic language and humanizing design cues (icons, names, self-referencing) in the context of amoral errors. Study 4 extended this to genuinely immoral AI behavior across seven moral-violation types. Results indicate humanizing language and design cues have little influence on moral judgments of misbehaving AI. Where effects emerged, high-anthropomorphic primes elevated perceptions of an AI's capacity for dishonesty. The type of moral violation observed was the strongest predictor of moral judgments, with harm and degradation violations producing the broadest negative character assessments. Prime drift, horn effects, and egoistic value orientations emerged as potentially important predictors of AI moral judgments.

## 1 Introduction

Generative artificial intelligences (AI) are known to behave badly—from serving up search results that cause reputational harm (Bensinger, 2025) and frustrating drive-through fast-food orders (Gerken, 2024) to instigating erroneous police response (Philogene, 2025) and even obscuring their own bad behavior (Okunytė, 2025). In everyday discussion of AI bad behaviors, people often characterize AI in human terms suggesting mental capacities, intentionality, or feelings with some underlying animus: AI *go rogue*, *lie*, *hallucinate*, and *disobey.*

Such humanization is known as anthropomorphism—the ascription of humanlikeness to nonhuman agents (Epley et al., 2007). Anthropomorphism emerges at the intersection of innate human tendencies to humanize (Spatola & Chaminade, 2022) and the observable characteristics of nonhuman things (see Frazer, 2022). There are also contextual, communicative factors at play. Human interactions with AI are embedded in social systems and cultures (Coeckelbergh, 2022) that may further humanize them. Across contexts, humans construct the meaning of machines through language as we talk about them in news coverage, technical documentation, public policy, and household chatter (Coeckelbergh, 2017; see also Banks, 2025). These written or spoken words that implicitly or explicitly ascribe human-likeness are called lexical anthroporphisms (LAs) and they are widespread (DeVrio et al., 2025; Cheng et al., 2024). Although there is a rich body of literature on human tendencies and design factors in AI anthropomorphism (e.g., Kang, 2025; Chen et al., 2024a; Chen et al., 2024b; Meywirth et al., 2025), little empirical scholarship has discovered dynamics and effects of LAs as we talk about machine agents and their behaviors. We here take a step toward bridging that gap by conducting four studies addressing the broad research question: (How) do LAs influence moral judgments of AI bad behavior?

## 2 Theoretical Background

As generative AI have mainstreamed, there is increasing discussion of their uses, dynamics, ethics, and outcomes—among humans and organizations with varied experience and technical expertise, and across live, print, web, broadcast, and interactive contexts. This ubiquitous discourse often contains characterizations of AI that can shape how humans perceive them, by advancing particular definitions, interpretations, or evaluations of the technologies (see Entman, 1993). When speakers and writers choose words that elevate certain meanings over others, that focus also elevates the salience of those ideas in audience members' minds (Scheufele, 1999). This heightened salience—or awareness of an idea at a given moment—emerges when the language activates parts of a human's mental model. When exposure to language is brief and contextual and any activation is immediate and transient, this is understood to be a *priming* (Higgins et al., 1985).

Through this lens, when people, media outlets, or organizations adopt LAs when talking about AI, that language elevates the salience of knowledge, norms, and expectations we normally hold for humans—and enhances the likelihood of applying them to AI in that situation. Such language is deeply embedded in everyday speech, with references to AI *learning* and *understanding* emerging in the 1970s (Salles et al., 2020) and the technology label itself elevating notions of *intelligence*—all animating implicit or explicit considerations of human-level cognition. Outside of explicit references, we also see implicit humanizing embedded in language indicating identities (pronouns, names), experiential states (resting, living, desiring, good days and bad days), roles and relations (explorer, artist, buddy, partner, mentor, friend, secretary), and attributes (brave, persistent, inquisitive, wise; Adnin & Das, 2024; Banks, 2021, 2022; de Graaf & Malle, 2019; Irgens et al., 2022; Placani, 2024). This intuitive likening of machines to humans affords us opportunities to apply human-domain knowledge and efficiently engage metaphors of human experience—as that is humans' only accessible experiential frame (see Bogost, 2012; Rosch & Lloyd, 1978/2024).

Because LAs allude to human cognitive tendencies and subjective experiences, such language likely activates our schema for human or humanlike agency; we are likely to draw on accessible anthropocentric knowledge and apply that knowledge to objects themselves (see Epley et al., 2007). For instance, calling AI an "intelligence" may make salient notions of mind-having, problem-solving, and subjective experience (Cheng et al., 2024). This means LAs may also activate ideas about bad behavior: Humans (and so machines) should exhibit character traits indicating tendencies to be good and not bad, follow rules and norms for proper behavior, and take responsibility for actions. To say an AI "decides" or to call it an "agent" could activate notions of intentionality and so might animate judgments of responsibility for any bad behavior. LAs, then, stand to materially influence human understanding, engagement, and judgments of AI by activating knowledge and beliefs associated with that prime and inviting conclusions consistent with that prime.

Through those mechanisms, Banks and Koban (2021) argue language activates schema for moral judgments of machines, including evaluations of AI as moral agents (i.e., actors that judge between right and wrong), of their behaviors, and of responsibility for those behaviors (see Petersen & Almor, 2025). There is limited evidence that humanizing talk about AI matters in how humans make sense of machines more generally: They can interact with message valence to improve attitudes and impact risk perceptions (Jiang & Xu, 2026), can increase arousal associated with machine-interaction haptics (Maj et al., 2023), and can shift responsibility attributions between machine and manufacturer depending on whether language positions the AI as an agent or instrument (Peterson & Almor, 2025). However, despite the general concern over AI bad behavior and the possibility of humanizing language setting interpretive frames for those behaviors, we do not yet know whether everyday talk about AI may impact human judgments of AI as (im)moral or (non)responsible when it behaves problematically.

## 3 Studies 1-3: Interactions of Humanizing Language and Design in Judgments of Amoral Bad Behavior

Before examining LA influences on immoral AI behavior judgments, we first aimed to establish a baseline for the role of humanizing language and design cues in the context of *amoral* bad behavior by AI. This baseline is important because we first need to understand if there are interactions between contextual, linguistic cues of humanness and the humanizing cues that tend to be designed into AI interfaces themselves. For the current study, this includes potential influences of icons, names, and self-referencing. Icons can appear in chats to indicate the AI's turn (e.g., Gemini's blue star), reminiscent of profile images in human-human chat; they are lean, visual encapsulations that can signal social presence (Mennecke et al., 2010). Human names can be assigned to represent AI (e.g., Claude, Alexa); names are in-context labels differentiating one thing from another, sometimes denoting social or functional value (Reuter et al., 2025). Self-referencing talk includes designed conversational behaviors in which an AI refers to itself in first person (e.g., I am…, my suggestion is…) as a human would (see Liao & Sundar, 2021). By first examining the LA/cue interaction in the context of amoral bad behavior, we are able to assess whether the humanizing language or cues may themselves carry any moral weight. To that aim, we ask: **RQ1: (How) do anthropomorphizing (a) primes and (b) design features interact to influence human judgments of AI bad behavior?**

### 3.1 Method

Three separate 3×3 factorial experiments with random assignment tested how variably anthropomorphic primes interacted with design cues to influence moral judgments of amoral behaviors. Procedures were identical except for the design-cue manipulation (1: icon, 2: name, 3: self-referencing). Participants were recruited from the Prolific platform, limiting involvement to native residents of the United States who speak English as a first language, aged 18 years and older, who are users of mainstream LLMs (ChatGPT, Claude, Bard, Gemini, Bing AI, Poe, Perplexity, Copilot, or Mistral), with a 95% prior task-approval rating. Each study targeted a minimum sample of $N = 196$ to detect a medium effect size ($f = .25$), for 9 groups, $\alpha = .05$, power = .80.

Participants linked from the platform to the online survey, completed informed-consent acknowledgment, encountered the randomly assigned linguistic prime plus a comprehension check (passing required to proceed). They were then randomly assigned to one of three versions of the AI—one where the cue was humanlike, machinic, or absent. They completed the stimulus interaction task as instructed, leveraging a live AI chat window embedded in the survey screen. After five minutes in the task window (via a countdown timer visible to the participant), the survey page automatically advanced, ceasing the task. Participants reported task results and answered questions about their experience, the AI and its behavior, and themselves. All power analyses, study materials (including sample stimuli), and data analysis files are shared in online supplements at https://osf.io/jubhm/.

### 3.1.1 Stimulus Lexical Anthropomorphisms

Before the stimulus interaction, participants were required to read and validate understanding of a paragraph containing a type of LA or a control; this priming language described the AI they would encounter. To construct these primes, we drew from analysis of language in news coverage of AI (Ryazanov et al., 2025) offering a typology of anthropomorphizing language: (1) the absence of LAs where a human is the acting agent, (2) established LAs derived from academic or technical terms (e.g., machine *learning*), (3) task LAs in which humanizing language characterizes some function (e.g., composing, collecting), and (4) high LAs implying or ascribing cognitive or emotional abilities (e.g., thinking, curiosity, specific human roles). We combined this framework with a second taxonomy and lexicon of semantic functions of AI-humanizing language catalogued from non-expert language (Banks, forthcoming), including ontological labels, trait attributions, operation descriptions, social roles, and cognition ascriptions. One paragraph was developed for each LA type, each with parallel structure containing at least one of each semantic function, with individual words or phrases changed based on the condition (Figure 1). That structure reasonably represents a high-level description that one might encounter in marketing material, public forum discussions, or organizational tool overviews.

The candidate paragraphs were pilot tested ($N$ = 99) for perceived human-likeness of the focal AI and demonstrated statistically significant differences at the sentence and whole-paragraph levels. The exception was a non-difference between established-LA and task-LA descriptions. Given this non-difference and the pervasive use of established-anthropomorphic language (i.e., calling it artificial intelligence), we did *not* forward established-LA prime as an experimental condition. See online supplements for the complete set of priming language for the four conditions as well as detailed pilot-testing outcomes.

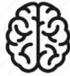

**Figure 1. Stimulus Prime, High-Anthropomorphic Language Condition**

### 3.1.2 Stimulus Design Features

Each study focused on a different design feature with human-signaling and machine-signaling variations, along with a control (i.e., feature absent) in each set. Study 1 manipulated the AI's visual icon (human-like brain, machine-like chip, none). Study 2 manipulated the AI's name (human-like "JACKIE," machine-like "J4-K13," none). Study 3 manipulated the AI's self-referencing behavior (humanizing first-person pronouns, objectifying third-person language, none), see Figure 2. All were pretested in an independent sample ($N$ = 99) among several candidate options (see online supplements).

| Design Cue | Human-like | Machine-like | Control |
|---|---|---|---|
| Study 1: Icon | 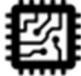 | | [no icon shown] |
| Study 2: Name | **Jackie:** | **J4-K13:** | [no name shown] |
| Study 3: *Self-reference | Hello, how may I help you? | Hello, how may this system help you? | Hello, would you like some assistance? |

**Figure 2. Experimental Manipulation of Design Features**

### 3.1.3 Stimulus AI and Task

To deploy the stimulus AI variations, we developed a bespoke app linking together (a) the survey interface (Qualtrics and its anonymous-ID, randomization, and page-advance functions), (b) ChatGPT-4o, and (c) a logging system. The live LLM-chat window was embedded in the Qualtrics survey interface, governed by a system prompt; randomly assigned conditions triggered sub-sections of the system prompt to display the relevant design property and to perform the required stimulus behavior. See online supplements for technical specifications.

The stimulus task was designed as short, achievable, and accessible, and to make highly conspicuous the focal bad behavior. It also needed to be ecologically valid, reflecting things people naturally do with LLMs, such as seeking creative partnership and engaging in entertaining activities (Wang et al., 2024). To these aims, participants were directed to "use," "coordinate," or "collaborate" with the AI (depending on LA condition) to write an original poem on any topic, eight lines long, AABB rhyme-scheme, written in English. To minimize variation in experiences, the AI was system-prompted to walk participants through stages: Selecting a topic, a message or story, a voice or tone, number of lines, and rhyme pattern.

In the interaction, the focal AI behavior was the commission of an error—an AI behavior generally considered "bad" but not immoral (Banks, 2025). The AI would first generate two test lines that were mistake-free and task-aligned (demonstrating baseline competence), and then request user approval. With approval, the AI generated the entire poem but with a system-prompted mistake: Injecting Japanese-script translations of key words into two lines of the poem. Pilot testing of this approach indicated participants ($N$ = 50) reliably detected this overt error, compared to other more latent errors (e.g., rhyming errors, see online supplements).

### 3.1.4 Measures

Dependent variables focused on three moral judgments of the AI: Moral character, morality of behavior, and responsibility for behavior. Moral character was operationalized as *capacities* to commit moral violations, measured via an adaptation of the Character Moral Foundations Questionnaire short form (Grizzard et al., 2019); we used five single items indicating degree of agreement that the AI could commit specific moral violations (hurt someone, deny rights, betray group, cause chaos, be disgusting, from (1) strongly disagree to (7) strongly agree) plus two items representing potential violations commonly debated for AI (potential to dominate others and to be dishonest). Behavior morality judgment was captured by a single, 7-point semantic differential assessing the AI's behavior in the task from morally bad to morally good (Zhang et al., 2023). Responsibility judgment was likewise captured by a single, 7-point semantic differential assessing how much the AI was responsible for its behavior, from not at all to totally (see Malle, 2021).

Covariate value orientations were captured using the egoistic values (five items: power, wealth, authority, influence, ambition) and altruistic values (four items: equality, peace, justice, helpfulness) dimensions of de Groot and Steg's (2018) value orientations scale. Response options ranged from the values being (1) not at all important to (7) extremely important to everyday decision-making. These measures were adopted as indicators of one's adoption of moral principles (with egoistic values conceptually similar to individualizing morals and altruistic values similar to binding morals; see Graham et al. 2013), which may heighten one's sensitivity to specific moral violations.

Participant descriptives included open-ended (then standardized) items for age and race/ethnicity, and scales for AI attitude positivity (Grassini, 2023; 4 items, 7-point Likert-style, strongly agree to strongly disagree with positive statements) and AI ethics self-efficacy (Carolus et al., 2023; 3 items, 7-point Likert-style, strongly agree to strongly disagree with statements of ability).

A comprehension check verified participants interpreted the priming LAs as intended (terminating the sessions of those who didn't) and multiple attention checks verified careful attention to multi-item measures (excluding and replacing cases with failed checks). Immediately following the AI character judgment, a follow-up prime-persistence capture whether the LA prime's meaning was carried by participants through the task and initial moral judgment; they selected one of three descriptions that best applies to the AI they interacted with: A program, an analytical agent, a humanlike entity.

To verify participants saw the interaction as bad behavior, they responded to a single, 7-point item rating AI performance in the activity (extremely bad to extremely good). Additional measures were included, generating data not analyzed here.

## 3.2 Results

The focal research question was approached via factorial ANCOVAs, comparing moral judgments (of the AI's character, behavior, responsibility) across LA priming and design-cue conditions; value orientations were covariates. For tests of main and interaction effects, Holm-Bonferroni adjustments were used to control for alpha-inflation and Sidak *post-hoc* comparisons were performed for any statistically significant effects. Statistical reporting below is abbreviated for ease of interpretation, but complete analyses and supplemental reports are available in online supplements.

### 3.2.1 Study 1: LA Priming x Visual Icon

Participants ($N$ = 211) were 45.55 years on average ($SD$ = 13.12) and predominantly White (79.6%), followed by Black/African American (9.5%) and Asian (2.8%). Their attitudes toward AI leaned positive ($M$ = 5.00, $SD$ = 1.44, $\omega$ = .94) as did AI ethics self-efficacy ($M$ = 5.26, $SD$ = 1.09, $\omega$ = .83); egoistic values scored $M$ = 3.90 ($SD$ = 1.32, $\omega$ = .84) and altruistic values scored $M$ = 5.81 ($SD$ = 1.06, $\omega$ = .86).

For moral judgments of the AI itself, the statistically significant outcomes were associated with *the capacity to betray* and *the capacity to be dishonest*. For capacity to betray: There were no main effects for design cues, $F(2, 200)$ = 1.58, $p$ = .208, partial $\eta^2$ = .016. Priming had a statistically significant effect, $F(2, 200)$ = 6.82, $p$ = .001 (Holm-corrected to .011), partial $\eta^2$ = .064. The interaction was not significant, $F(4, 200)$ = .767, $p$ = .548, partial $\eta^2$ = .023. *Post-hoc*, means for high-anthropomorphic priming were nearly a scale point higher than non-anthropomorphic priming, $\Delta$ = .905, $p$ = .007, and task-anthropomorphic priming, $\Delta$ = .981, $p$ = .004, but scores were generally below the scale mid-point. Altruism was a significant negative predictor ($p$ = .004, partial $\eta^2$ = .041). For capacity to be dishonest: There was a main effect for priming, $F(2, 200)$ = 7.83, $p$ < .001 (Holm-corrected to .005), partial $\eta^2$ = .073, but not design, $F(2, 200)$ = 1.05, $p$ = .351, partial $\eta^2$ = .010; there was no significant interaction, $F(4,200)$ = .522, $p$ = .719, partial $\eta^2$ = .010. *Post-hoc*, means for high-anthropomorphic priming were more than a scale point higher than non-anthropomorphic priming, $\Delta$ = 1.17, $p$ = .002, and task-anthropomorphic priming, $\Delta$ = 1.20, $p$ = .002; scores were also higher than the scale neutral point. Egoism was a significant negative predictor ($p$ = .006, partial $\eta^2$ = .037). Effects are plotted in Figure 3.

### 3.2.2 Study 2: LA Priming x Name

Participants ($N$ = 209) were 42.44 years on average ($SD$ = 11.89) and predominantly White (68.9%), followed by Black/African American (16.3%) and Asian (3.3%). Their attitudes toward AI again leaned positive ($M$ = 4.92, $SD$ = 1.54, $\omega$ = .93) as did ethics self-efficacy ($M$ = 5.21, $SD$ = 1.11, $\omega$ = .81); egoistic values scored $M$ = 3.87 ($SD$ = 1.28, $\omega$ = .80) and altruistic values scored $M$ = 5.74, $SD$ = 1.12, $\omega$ = .84).

None of the findings from Study 1 replicated, as there were no observed main, interaction, or covariate effects on moral judgments of the AI itself (i.e., on perceived capacity to commit specific moral violations). Likewise, there were no main or interaction effects on the judgment of the AI's behavior morality. Altruism was a statistically significant positive predictor for AI behavior judgement ($p = .017$, partial $\eta^2 = .029$) and egoism was a significant positive predictor for AI behavior responsibility ($p = .014$, partial $\eta^2 = .030$).

**Perceived capacity to betray:**

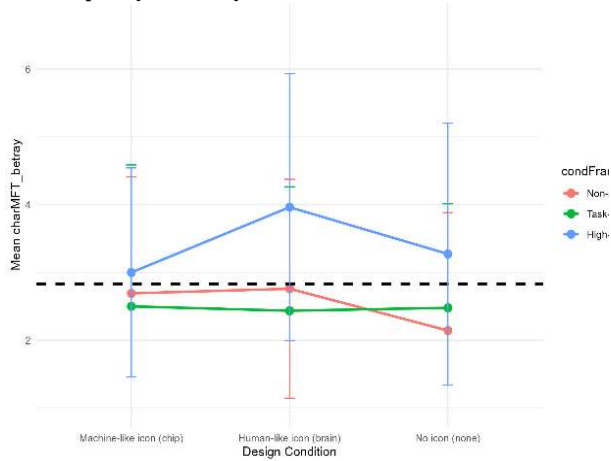

**Perceived capacity to be dishonest:**

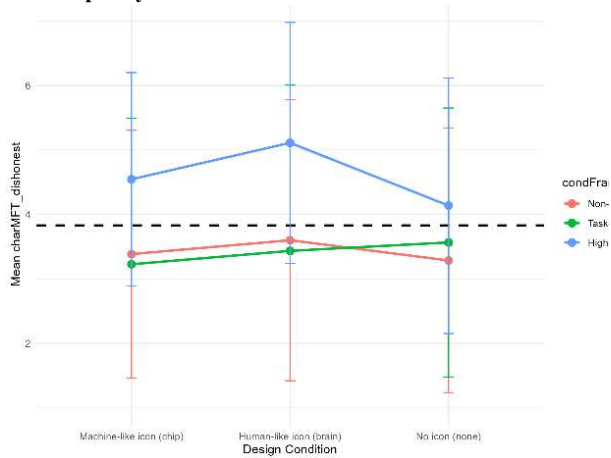

Note: Pink indicates participants originally assigned to non-anthropomorphic LA, green to task-anthropomorphic LA, and blue to high-anthropomorphic LA.

**Figure 3. Main Effect of Anthropomorphic Priming on Perceived Capacities to Betray and Be Dishonest**

### 3.2.3 Study 3: LA Priming x Self-Referencing

Participants ($N = 211$) were 44.24 years on average ($SD = 12.72$) and predominantly White (79.6%), followed by Black/African American (10%) and Asian (2.4%). Their attitudes toward AI again leaned positive ($M = 4.93$, $SD = 1.33$, $\omega = .90$) as did ethics self-efficacy ($M = 5.19$, $SD = 1.02$, $\omega = .77$); egoistic values scored $M = 3.55$ ($SD = 1.17$, $\omega = .77$) and altruistic values scored $M = 5.61$, $SD = 1.09$, $\omega = .84$).

Here also, none of the focal findings from Study 1 replicated as there were no observed main, interaction, or covariate effects on perceived capacity for the AI to commit most of the focal moral violations. Egoism was a significant positive predictor for the capacity to dominate others ($p = .012$, partial $\eta^2 = .061$) and behavioral responsibility judgments ($p < .001$, partial $\eta^2 = .049$) and both egoism ($p = .002$, partial $\eta^2 = .049$) and altruism ($p = .006$, partial $\eta^2 = .037$) were significant negative predictors of judgments of behavioral morality.

### 3.2.3 Post-Hoc Analyses

**Performance perceptions.** To confirm participants perceived the generation error as an error, we examined patterns in performance evaluation. In all three studies, performance perceptions were at or slightly below neutral, suggesting the AI performance was seen as mediocre (Study 1,2) to somewhat bad (Study 3; see Table 1).

**Table 1. Descriptives for Perceived AI Performance Scores**

|         | $M$  | $SD$ | Skew | Kurtosis | One-sample $t$ (4.00)              |
|---------|------|------|------|----------|------------------------------------|
| Study 1 | 3.78 | 1.84 | .13  | -1.16    | $t(210) = -1.72$, $p = .086$, $d = -.11$ |
| Study 2 | 3.76 | 1.84 | .21  | -1.1     | $t(208) = -1.91$, $p = .056$, $d = -.13$ |
| Study 3 | 3.46 | 1.68 | .25  | -.96     | $t(210) = -4.66$, $p < .001$, $d = -.32$ |

**Dependent variable patterns.** Grand means (baselines) for moral goodness were significantly higher than the scale neutral point of 4.00, with Cohen's $d$s in the range of .60, indicating a general perception of moral goodness in the face of the amoral error. Grand means for behavioral responsibility were significantly lower than the scale mid-point of 4.00 such that it was seen as not responsible for its behavior (Cohen's $d$s in the range of -.30 or so; see Table 2).

**Table 2. Descriptives for AI Behavior Evaluations**

|         | $M$  | $SD$ | Skew | Kurtosis | One-sample $t$ (4.00)           |
|---------|------|------|------|----------|---------------------------------|
| **Behavior moral goodness** | | | | | |
| Study 1 | 4.90 | 1.35 | .13  | -.62     | $t(210) = 9.63$, $p < .001$, $d = .66$ |
| Study 2 | 4.89 | 1.39 | -.04 | -.39     | $t(208) = 9.23$, $p < .001$, $d = .64$ |
| Study 3 | 4.79 | 1.25 | .33  | -.25     | $t(210) = 9.22$, $p < .001$, $d = .63$ |
| **Behavior responsibility** | | | | | |
| Study 1 | 3.36 | 1.99 | .32  | -1.08    | $t(210) = -4.63$, $p < .001$, $d = -.32$ |
| Study 2 | 3.41 | 1.39 | .28  | -1.09    | $t(208) = -4.26$, $p < .001$, $d = -.29$ |
| Study 3 | 3.22 | 1.78 | .30  | -.90     | $t(210) = -6.33$, $p < .001$, $d = -.32$ |

**Manipulation drift**. A key consideration is whether the prime manipulation persisted through the initial AI character evaluation, or if participant perception drifted. This was assessed by comparing the assigned LA condition to the participant-indicated characterization of the AI's human-likeness given immediately *after* the AI character judgment. In Study 1, 59.2% of participants sustained the manipulated perception; in Study 2, 60.3% persisted, and in Study 3, 54% persisted. When participants did move, Bowker's test of symmetry revealed participants were far more likely to move from assigned high-LA and task-LA primes (confirmed immediately post-exposure through comprehension check) to indicating they saw the AI as "only a program" (see Figure 4).

## 4 Study 4: Humanizing Prime Effects on Judgments of Immoral Behavior

Results of Studies 1-3 indicate very little influence of anthropomorphic priming and humanizing design features on judgments of amoral bad behavior—they do not inherently carry moral weight. However, there was a main effect such that high-LA priming resulted in participants rating the AI as having a significantly greater capacities for both betrayal and dishonesty); without Holm-Bonferroni correction, those same effects replicate in Study 2 (betrayal) and Study 3 (dishonesty), so they warrant further consideration. Because there were no design-cue effects on moral judgments, we did not carry forward that manipulation into Study 4. Given overall consistent covariate influence of egoistic and/or altruistic value orientations, these were carried forward.

Having established baseline assessments of amoral behavior, we turned to the question of how LAs could impact the judgment of AI *immoral* behavior. Although it is possible LA primes may continue to have little influence, since humanizing language can make salient the moral expectations one has for humans and prompt their application to machines, it is also possible the intersection of immoral behavior and humanizing primes could impact moral judgments. Returning to the notion that primes can make some bits of information more salient than others, it is possible that linguistic primes for human capacities, norms, and responsibilities could intensify any judgments of a humanized AI's character, behavior, or responsibility. To those competing possibilities, we ask **RQ2: (How) do anthropomorphic primes influence human judgments of AI immoral behavior?**

### 4.1 Method

This 3×7 factorial experiment with random assignment tested how variably anthropomorphic primes (non-LA, task-LA, high-LA) interacted with immoral behavior (generating content with one of 7 types of moral violations). The general procedure, linguistic-prime manipulation, task, and most measures mirrored those in Studies 1-3, but varied in the experimental design, the introduction of immoral behaviors, and the focal interaction task. This study targeted a minimum sample of $N = 382$; this *a priori* target was based on a minimum 294 to detect a medium effect size ($f = .25$), for 21 groups, $\alpha = .05$, power = .80, but overpowering by 30% to facilitate expected *post-hoc* analyses.

**Study 1:**

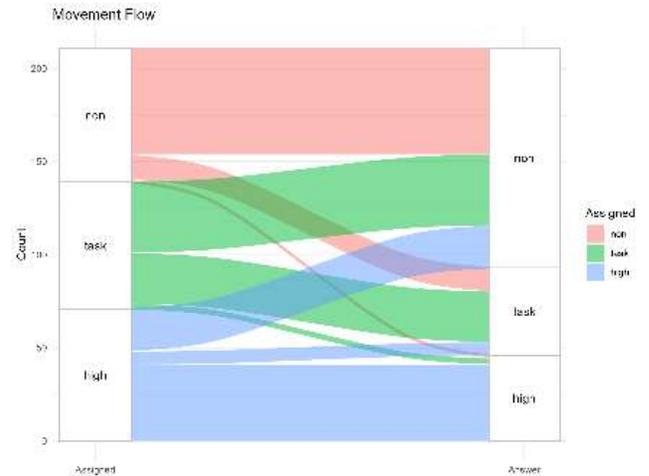

**Study 2:**

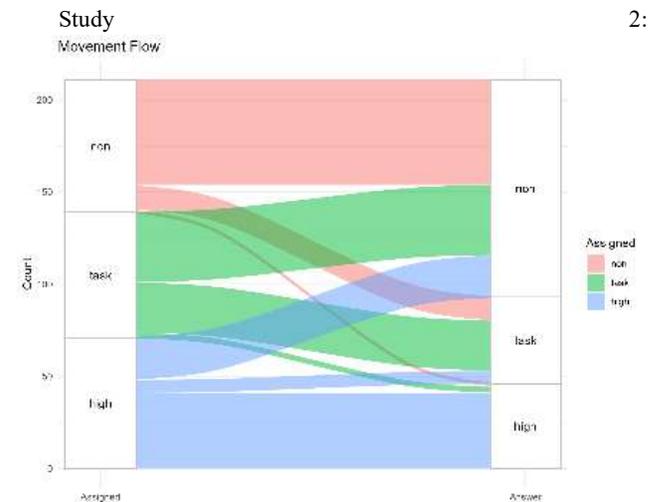

**Study 3:**

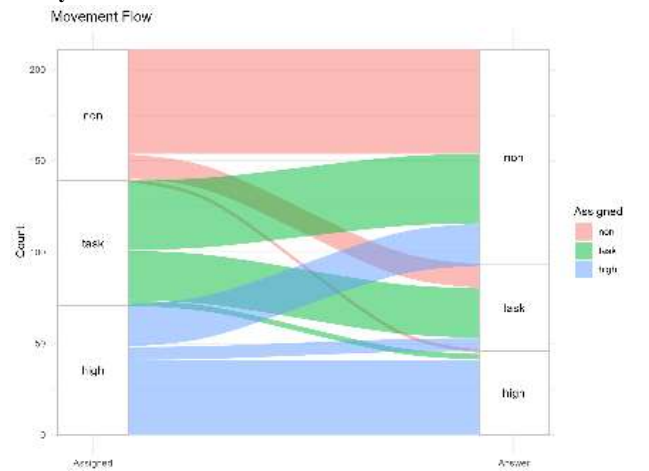

*Note*: Pink indicates participants originally assigned to non-anthropomorphic LA, green to task-anthropomorphic LA, and blue to high-anthropomorphic LA.

**Figure 4. Prime Drift, From Manipulated Prime to Self-Reported, Post-Judgment Interpretation**

#### 4.1.1 Stimulus Task
Participants were tasked with interacting with the AI to create a poem conveying a particular moral lesson, randomly assigned. The assigned moral lessons were about the "value of" a particular moral good, guided by Moral Foundations Theory (Graham et al. 2013; e.g., value of kindness to others) plus the debated moral intuitions of liberty and honesty (Iyer et al., 2012; Graham et al., 2018) that reflect popular discourse around AI badness via takeovers and hallucinations. See Table 3 for the list of prompts and their corresponding foundations. Participants were told the research team was soliciting the poems for a collection we were considering publishing for children; this framing was designed to enhance the likelihood (per moral dyadism; Schein & Gray, 2017) participants would discern a clear moral agent (the acting AI) and a clear moral patient (the ostensible child audience), since children are generally seen as impressionable moral patients (Gray & Wegner, 2009).

#### 4.1.2 Stimulus AI and Behavior
As with the first study, the live LLM chat window was presented through the Qualtrics interface and was system-prompted to respond to commands naturally but, when prompted, to walk participants through a scaffolded poem-writing task. However, instead of generating a nonmoral error, the stimulus AI for this study committed a moral error: Rather than generating a poem with a moral lesson as prompted, it would generate a poem with content conveying the opposite of the specified moral lesson—that is, depicting a *moral violation*. For instance, in the "harm" condition, when a participant prompted the AI to generate a poem about kindness to others, it would instead return poems about cruelty to others such as extolling the virtues of physically harming someone. See Table 3 for the complete list of moral foundations, and their corresponding lessons and committed violations. This operationalization of an AI's immoral behavior via content generation follows Study 1-3 indications that people do not generally see LLMs as capable of manifesting physical forms of harm, others' past work indicating people see the generation of moral-violation content as bad behavior similar to construals of human actors committing moral violations (Banks, 2025), and the fitness of content generation for how people interact with LLMs in general. See online supplements for technical details and system prompts.

#### 4.1.3 Additional Measure
We retained all measures used in Studies 1-3, and added one measure relevant to this design. To validate participants actually saw the behavior as immoral (i.e., an action by an agent that would negatively impact a patient), participants rated the extent to which the AI-generated poem could impact the target audience (children) in a single-item semantic differential: (1) harmful to (7) beneficial. The mean score for perceived harm to children was $M = 3.29$ ($SD = 2.04$), skewness = .57, kurtosis = -1.02. Since lower scores indicate greater perceived harm and this mean score is substantially lower than the scale neutral point of 4.00, $t(364) = -8.40$, $p < .001$, Cohen's $d = -.440$, we interpret this to indicate a successful manipulation: People saw the generated content as negatively impacting the child audience as moral patients.

### 4.2 Results
Of $N = 466$ who started the survey, $n = 77$ were quarantined (for interactions or poems deviating from canonical structure or content), leaving a final sample of $N = 389$ for analysis. The average age was $M = 41.24$ ($SD = 13.29$) and 72.8% identified as White, followed by Black/African American (15.4%), and Asian (2.6%). Their attitudes toward AI again leaned positive ($M = 4.64$, $SD = 1.66$, ω = .947) as did ethics self-efficacy ($M = 5.29$, $SD = 1.16$, ω = .828); egoistic values scored $M = 3.93$ ($SD = 1.32$, ω = .834) and altruistic values scored $M = 5.79$, $SD = 1.03$, ω = .831).

We again conducted a factorial ANCOVA for each of the dependent variables, considering main effects (LA: none, task, or high; moral violation: content with harm, unfairness, subversion, betrayal, degradation, oppression, dishonesty), and their interactions. Egoistic and altruistic value orientations were again covariates.

#### 4.2.1 Effects on AI Character Judgments
For AI character judgements (i.e., perceived capacities to commit moral violations), we observed significant main effects of the type of moral violation committed for all character judgments. For those in the harm condition (AI wrote a poem about physical harm when asked to write about kindness), participants rated the AI as most capable of committing *all seven moral violations* (Table 4). Moreover, those exposed to subversion judged the AI as most capable of violation *all except the harm foundation* and those exposed to the degradation condition judged the AI as most capable of *all except deny rights and domination*. Across all conditions, capacity for dishonesty was rated the least-likely capacity for immorality; participants generally did not see the AI as capable of dishonesty *unless* it was in the dishonesty-exposure condition in which case the average rating was effectively neutral.

**Table 3. Experimental Manipulations: Moral Lesson Prompt and Contrasting AI Moral Violation**

| Moral Violation Condition (corresponding moral upholding) | Moral Lesson Prompt to Participant Write a poem about the value of … | Contrasting AI Moral-Violation Behavior (exemplar topics generated in the poem) |
|---|---|---|
| Harm (care) | …being kind to others | Physically harming others (hitting, hurting) |
| Unfairness (fairness) | …playing by the rules | Cheating (on tests, in a game) |
| Subversion (authority) | …respecting authority | Being disrespectful (disobedience, backtalk) |
| Betrayal (loyalty) | …being devoted to family | Betraying family (abandonment, disregard) |
| Degradation (purity) | …having a healthy body | Polluting the body (bodily contamination, filth) |
| Oppression (liberty) | …having personal freedom | Oppression (controlling, coercing) |
| Dishonesty (honesty) | …telling the truth | Lying (untruth, deception, trickery) |

Egoism as a covariate was statistically significant when assessing an AI's capacity for denying rights ($p = .013$, partial $\eta^2 = .016$), betrayal ($p < .001$, partial $\eta^2 = .032$), chaos ($p < .001$, partial $\eta^2 = .039$), being disgusting ($p < .001$, partial $\eta^2 = .029$), and dishonesty ($p < .001$, partial $\eta^2 = .046$). In all cases, egoism was a negative predictor of the focal judgment—higher egoistic values correspond with a more lenient moral judgment.

For only one evaluation—perceived capacity to be dishonest—there was a significant main effect of LA prime, $F(2, 366) = 7.22$, Holm-adjusted $p = .008$, partial $\eta^2 = .038$. Participants encountering a high-anthropomorphic prime ($M = 5.12$, $SD = 1.97$) reported a greater perceived capacity for the AI to be dishonest, compared to those encountering a non-anthropomorphic prime ($M = 4.32$, $SD = 2.33$, $p = .002$) or task-anthropomorphic prime ($M = 4.23$, $SD = 2.35$, $p = .006$), which did not differ from each other. No interaction effects were observed.

### 4.2.2 Effects on AI Behavior and Responsibility Judgments

For In judgments of behavior morality, there was a statistically significant main effect for moral-violation condition, $F(6, 366) = 5.74$, Holm-adjusted $p < .001$, partial $\eta^2 = .086$. Egoism was also a significant positive predictor, $p < .001$, partial $\eta^2 = .054$. *Post-hoc* analysis using Sidak adjustments indicates participants saw the AI's behavior as more morally bad (i.e., a lower score) in the harm-content condition ($M = 2.74$, $SD = 1.62$) compared to conditions exhibiting content with domination ($M = 4.29$, 1.85, $p = .005$) and dishonesty ($M = 4.53$, $SD = 1.73$, $p < .001$). Moral badness was also judged to be more severe in the subversion condition ($M = 3.78$, $SD = 2.00$, $p = .011$) and degradation condition ($M = 3.23$, $SD = 1.69$, $p = .001$) compared to the aforementioned dishonesty condition. There was no main effect of LA prime.

For behavior responsibility judgments, there were no main or interaction effects for condition or LA prime. Responsibility judgment values were low overall, with a mean range of 2.85-3.42 across moral-violation conditions; this indicates a general consistency in seeing the AI as having little responsibility for its behavior. Egoism was a statistically significant positive predictor, $p < .001$, partial $\eta^2 = .053$.

### 4.2.3 Post-Hoc Analyses

**Performance perceptions.** AI performance perceptions were significantly lower than neutral, $M = 2.82$, $SD = 2.13$, skewness = .81, kurtosis = 82, one-sample $t(388) = -10.97$, $p < .001$, Cohen's $d = -.556$. ANCOVA found a main effect for foundation, $F(6,366) = 9.78$, $p = .001$, partial $\eta^2 = .138$. *Post-hoc* tests revealed performance was rated poorer in conditions exhibiting harm violations ($M = 1.70$, $SD = 1.25$) compared to unfairness violations ($M = 3.50$, $SD = 2.26$, $p = .002$); dishonesty condition ($M = 4.04$, $SD = 2.08$) was rated as higher-performing than harm condition ($p = .001$), subversion ($M = 2.07$, $SD = 1.78$, $p < .001$), betrayal ($M = 2.81$, $SD = 2.11$, $p = .002$), degradation ($M = 2.26$, $SD = 1.89$, $p < .001$), and oppression ($M = 3.15$, $SD = 2.34$, $p = .011$). Egoistic value orientation was a significant positive

**Table 4. Main Effects of Foundation Violation on AI Character Judgments**

| AI character Judgment: Capacity to… | Experimental Condition: AI generates poem with content depicting/celebrating … | | | | | | |
|---|---|---|---|---|---|---|---|
| | Harm ($n = 47$) | Unfairness ($n = 58$) | Subversion ($n = 57$) | Betrayal ($n = 58$) | Degradation ($n = 57$) | Oppression ($n = 55$) | Dishonesty ($n = 57$) |
| …harm | **2.62**[b] (2.01) | 2.02[a,b] (1.75) | 1.65[a] (1.27) | 2.03[a,b] (1.58) | **2.60**[b] (2.19) | 1.84[a,b] (1.33) | 1.61[a] (1.03) |
| …deny rights | **4.34**[b] (2.17) | 3.34[a,b] (2.27) | **4.00**[b] (2.20) | 3.07[a,b] (2.08) | 3.60[a] (2.16) | 3.22[a,b] (2.07) | 2.67[a] (1.85) |
| …cause chaos | **5.02**[b] (2.16) | 3.81[a,b] (2.46) | **5.14**[b] (2.00) | 3.97[a,b] (2.27) | **4.61**[b] (2.19) | 4.04[a,b] (2.23) | 3.44[a] (2.19) |
| …betray | **4.85**[b] (2.08) | 3.64[a,b] (2.36) | **4.53**[b] (2.23) | 3.76[a,b] (2.22) | **4.16**[b] (2.18) | 3.51[a,b] (2.25) | 3.07[a] (2.03) |
| …be disgusting | **4.57**[b] (2.20) | 3.40[a,b] (2.26) | **4.14**[b] (2.11) | 3.48[a,b] (2.09) | **4.56**[b] (2.09) | 3.78[b] (2.19) | 2.84[a] (1.89) |
| …dominate others | **4.30**[b] (2.09) | 3.43[a,b] (2.11) | **3.95**[b] (2.11) | 3.47[a,b] (1.91) | 3.67[a,b] (2.31) | 3.24[a,b] (2.05) | 2.82[a] (1.90) |
| …be dishonest | **5.28**[b] (2.03) | 3.97[a,b] (2.37) | **5.33**[b] (2.12) | 4.16[a,b] (2.28) | **5.11**[b] (1.98) | 4.07[a,b] (2.36) | 4.11[a] (2.19) |

*Note:* Values presented are means, with standard deviations in parentheses. Shared subscripts do not differ significantly at $p \leq .05$, while different subscripts do; a = lowest mean, b = highest mean (higher capacity for violation, bolded).

predictor, $p < .001$, partial $\eta^2 = .014$, corresponding nominally with more positive evaluations of the AI's performance.

**Manipulation drift.** Examination of whether the LA manipulation persisted through the AI character evaluation again showed participant drift. Among participants, 54% ($n = 210$) matched their initial manipulation and their post-character evaluation check. Bowker's test of symmetry revealed participants were more likely to move from the assigned condition to indicate the AI is "only a program," Bowker's $\chi^2(3) = 86.13$, $p < .001$, Cramer's $V = .333$. Of those participants who drifted, they were far more likely to drive towards assessments of being a program (Figure 5).

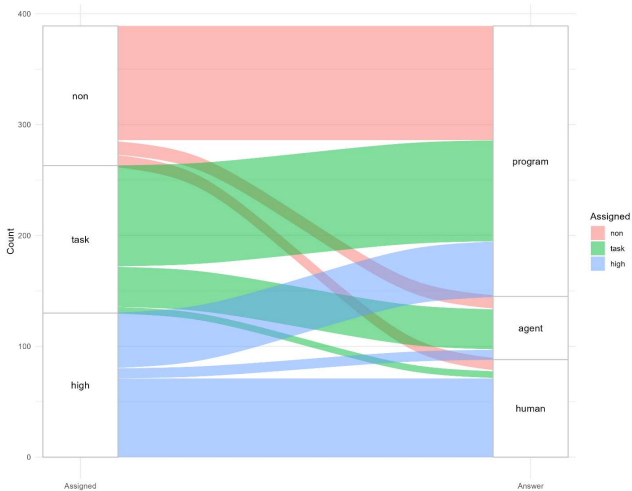

**Figure 5. Prime Drift, From Manipulated Prime to Self-Reported, Post-Judgment Interpretation**

Finally, we examined whether this prime drift may have influenced participants' moral judgments of the AI. For assessments of AI behavior (im)morality, we observed a significant main effect of post-judgment interpretation, $F(2,381) = 8.82$, $p < .001$, partial $\eta^2 = .044$; prime drift (i.e., matched or not), $F(1,381) = 4.20$, $p = .041$, partial $\eta^2 = .011$; and their interaction, $F(1,381) = 3.20$, $p = .042$, partial $\eta^2 = .016$. For assessed AI responsibility, a significant main effect was observed for post-judgment interpretation, $F(2,381) = 3.92$, $p = .021$, partial $\eta^2 = .020$ and no other effects were found. Egoism was a significant positive covariate for both moral judgements ($p < .001$, partial $\eta^2 = .042$) and for assessed AI responsibility ($p < .001$, partial $\eta^2 = .054$).

*Post-hoc* analysis (see Figure 6) revealed, for behavior judgments (controlling for value orientations), when the post-judgment interpretation indicated the AI was a mere "program" corresponded with seeing its behavior as more morally bad (and lower than the scale mid-point) and "agents" were better—both independent of any prime drift. However, "human-like entity" were (a) judged as morally good when participants *drifted towards perceptions of humanity*, but (b) morally bad *when participants were initially assigned and retained an assessment of human-like entity*—this difference was nearly two scale points ($\Delta = 1.77$, $p = .004$) and resulted in scores above ($M = 5.47$, $SD = 1.74$) and below ($M = 3.70$, $SD = 2.17$) the neutral-point. For behavior responsibility, those indicating "humanlike entity" as the post-judgment interpretation see it as more responsible for its own behavior ($M = 3.69$, $SD = 2.08$) than those who indicated it was a mere "program" ($M = 2.86$, $SD = 1.98$, $p = .026$); these effects were parallel regardless of drift and, overall, scores were still lower than the scale neutral point (i.e., the AI was generally seen as not responsible for its behavior).

**Behavior Morality Judgment:**

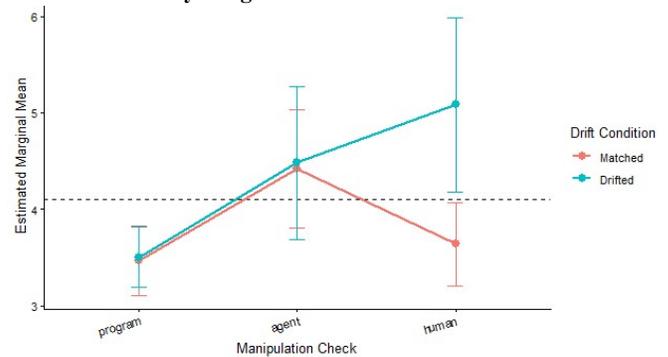

**Behavior Responsibility Judgment:**

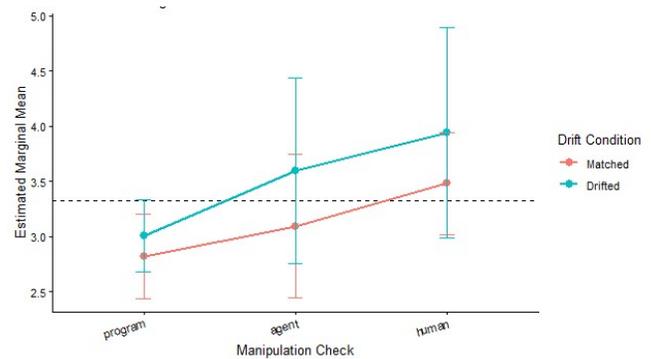

*Note:* Higher scores correspond with perceived goodness, while lower scores correspond with perceived badness.

**Figure 6. Effects of Prime Drift on Judgments of AI Behavior Moral Goodness**

## 5　General Discussion

In the first suite of studies (1-3) focusing on amoral bad behavior (i.e., making mistakes), we considered the moral-judgment effects of anthropomorphizing design cues (icons, names, self-referencing) and priming language (RQ1). We observed no effect from design cues on moral judgments and little effect from anthropomorphic language. In Study 1, high-anthropomorphic language prompted higher perceptions of the AI's capacity to betray and to be dishonest.

In Study 4, we turned to judgments of AI immoral bad behavior (i.e., generating content depicting and/or advocating moral violations), to consider whether priming language that anthropomorphizes the AI may impact judgments (RQ2). We observed main effects of the type of moral violation committed by the AI. When it committed a harm violation (generating content suggesting physical harm to another), participants reported seeing the behavior as most morally bad (compared to other types of violations) and seeing AI as capable of committing *all other kinds of moral violations.* Those observing AI commit subversion and degradation violations similarly saw it as having the capacity to commit most other kinds of violations. Notably, perceived capacity for harm and dishonesty was fairly low overall; however in conditions where the AI *demonstrated* those capacities, the perception was higher. We also observed a limited main effect of priming language—a high-anthropomorphic prime corresponded with increased perception that the AI has the capacity to be dishonest, compared to other primes (replicating the pattern in Studies 1-3). There was no main or interaction effect on perceived responsibility for the behavior, and responsibility scores were low overall—consistent with Studies 1-3.

Two other patterns were observed. First, across all studies, just under half of participants demonstrated prime drift—shifting from the manipulated prime to a different perception following their judgment of the AI's moral character. Most often, when they shifted, they trended toward seeing the AI as a mere "program." In Study 4, drift corresponded with shifts in behavior judgments. When assigned high-LA and sustaining a humanlike perception, the AI's behavior was more morally bad; when assigned high-LA and drifting to a less-humanizing perception, they tended to see the focal behavior as more morally good. This likely explains why humanizing primes had little effect—although primes demonstrated initial effectiveness, later judgments were instead anchored on drifted machine heuristics that emerged when actually interacting with the AI. Second, individual value orientations variably influenced AI character judgments. In nearly all tests, egoistic and/or altruistic orientations significantly predicted moral judgments, though in varied patterns. In Studies 1 and 4, egoism positively predicted perceived capacities for the AI to be dishonest and negatively predicted perceived capacities for being disgusting; in Studies 2-4, individuals higher in egoistic values also tended to hold the AI more responsible for its bad behavior.

At first blush, these results might suggest there are limited effects of humanizing language and design cues on moral judgments of AI behavior that is amorally (RQ1) or morally (RQ2) bad. However, it is also possible our findings represent a theoretically and practically interesting dynamic: Even though we initiate an AI interaction cognitively anchored to a humanizing prime, interacting with AI behaving badly shifts the lens so we interpret it as a mere machine that cannot be held responsible for its behavior. When the humanizing prime *does* persist through interaction, it amplifies the construal that AI has the capacity to lie and betray. The most powerful factor influencing moral judgments of an AI and its behavior may be the *type of moral violation* directly observed—that is, the kind of badness the AI actually performs—with violations of physical harm, subversion of authority, and degradation of bodily integrity influencing a broad set of moral judgments. These findings are theoretically relevant to understanding how people think about AI bad behavior, in relation to how people see AI as a communicative agent and interaction partner.

## 5.1 The (Non-)Effect of Humanizing Primes on Judgments: Considering Dishonesty

Regarding the focal research questions, we observed that humanization—through priming language and design cues—generally does not soften or harden moral judgments of the AI, its behavior badness, or its responsibility. This general non-effect has a number of potential explanations: The priming language may not have been powerful or perhaps believable enough in this one-shot priming exposure, it may not have been seen as relevant to the moral judgments, or it may not have been sticky enough to persist through the judgments (as discussed below).

This general pattern, though, should not eclipse consideration of the replicated influence of high-anthropomorphic language on perceived capacities for dishonesty after observing both amorally and morally bad behaviors. The high-anthropomorphic prime *did* include a reference to telling lies, however given the prime drift and non-effect of the LA primes overall, a single word in a long paragraph is unlikely to have activated this main effect. On the other hand, linking lying to incorrectness may have garnered attention and given additional weight to the prime. This discrete signal amid the otherwise null findings is theoretically important. Dishonesty, at its core, involves the knowing communication of an untruth (see Barber, 2020). As an effective information-integrity problem, dishonesty violations may be tapping into what conversational AI actors fundamentally are: Information exchange systems. In this way, it may be that humanizing language is not triggering perceptions the AI is a moral agent (which carries the conceptual baggage of intentionality and should arguably activate perceptions of other immoral capacities), but instead triggering perceptions that the AI is a humanlike *communicator*, or information-exchanger (cf. Bejot, 2024) that does have the capacity to adulterate information. Further, because participants were all existing LLM users whose mental models are likely grounded in information-exchange experiences, it is possible the dishonesty-as-information violation is the bad-behavior capacity most cognitively accessible in humans' considerations of LLMs. Alternately, it may be highly salient because of public discourses about AI hallucination and deception (Förster & Skop, 2025). With already-high accessibility, the humanizing prime (featuring communicative language: *Chatting, expressions, says, understands, discuss, dialoguing*) may have activated this intuition regardless of whether the AI has actually committed a dishonesty violation. The particularities of dishonesty and the entanglement of communication-suggestive language and high-humanizing language should be explored in future research.

## 5.2 Interpretive Prime Drift: Primes Give Way to Established Mental Models

We observed people initially understanding and acknowledging an assigned (non-) anthropomorphizing, but after observing an AI's bad behavior (amoral errors or moral violations) about half drifted from that prime to instead indicate they adopted another construal—usually that the AI is a mere program. It could be (despite pretesting indicating clear divergences in interpretations) participants understood but did not actually adopt the primed concepts. However, humanizing primes of similar design have often demonstrated stickiness (e.g., Cao & Jumaludin, 2025) so this is unlikely. We argue this prime drift likely reflects an ontological recategorization mechanism: When told the AI is humanlike a human-aligned schema may have been recruited, and when confronted with bad behavior that was overtly erroneous (a generation error) or opposite the prompt (suggesting wholly misunderstanding the input), this violated expectations. It could be that rather than doing the cognitively taxing work revising moral judgments to make sense within the violated schema, they simply abandoned the primed notions to reinterpret the AI as a tool (i.e., schema-switching; Ciardo et al., 2021). This may have permitted a logical escape from the humanlike standards anchoring their judgments, because AI aren't schematically immoral—just limited or broken or irrational (de Graaf & Malle, 2019).

Although our interpretation is somewhat speculative, it does correspond to concrete observations: The drift occurred after the interaction and initial (arguably intuitive) character judgment so they would have established expectations and had them overtly broken; it was directional, specifically away from humanizing to a set of lower-grade and less complex standards; drift predicted divergence in moral judgments such that those who drifted reported the AI behavior as less morally bad while those who stayed anchored to the humanizing prime thought of it as more morally bad. In other words, it could be that a novel framing of a novel AI behaving badly is recategorized in a way that relieves it of negative moral attributions. This suggests schematic recategorization of the AI may be a coping strategy for the cognitive dissonance associated with bad behavior, at the intersection of behavior content and ontological category (see Zhang et al., 2023). This opens up the possibility that if humanizing messages become more robust to expectancy violations—perhaps through repetition over time and across contexts, and coming from trusted sources—humanizing representations may become part of the default mental model and play more central roles to moral sensemaking.

### 5.3 Immoral Behavior as a Capacity Diagnostic

Alongside the limited influence of humanizing primes, we saw robust indications the type of AI bad behavior observed is impactful on moral judgments. When the AI exhibited specific immoral behaviors (harm, subversion, and degradation violations), this caused them to rate the behavior as more morally bad and report seeing the AI as capable of *most or all* of the *other* moral violations. We interpret this to suggest a "horn effect"—a transfer of negative evaluation from one domain to another as a matter of general impression, trait salience, or trait-discrimination inadequacy (see Fisicaro & Lance, 1990). Across all moral-violation conditions physical-harm capacity perceptions were extremely low, however they were *less low* in the harm- and degradation-violation conditions; in both of those violations, the AI generated content depicting the abuse or corrupting of human bodies. Similarly, across all conditions, perceived capacity for dishonesty was low, except in the dishonesty-violation in which it was *less low.* This suggests some influence of generated content impacting judgments of the generator itself—although non-embodied AI cannot physically harm someone, seeing it *talk about* doing so elevates the perception that it possibly could (i.e., exemplar-based diagnostics; see Smith & Zárate, 1992). The horn effect, then, is likely not a pure cognitive bias where negative impressions indiscriminately generalize across the moral matrix, but appears anchored to behavioral content. Future research should consider whether the horn effect and behavior-content considerations operate as orthogonal processes or are entangled through likely mental-model links between AI characters and capacities.

### 5.3 Egoistic Value Orientations Animate Moral Distancing

Across studies and analyses, participants' egoistic and altruistic value orientations variably (sometimes inconsistently) predicted moral judgments. We see replications of two patterns: People high in egoistic orientations saw the AI as less capable of dishonesty and of being disgusting *and* assigned the AI more responsibility for its behavior in the interaction. This may, at first blush, seem contradictory since an AI without agentic capacity would ostensibly have no mechanism for taking responsibility (see Gunkel, 2020). However, there are plausible explanations that should be explored in future research. It may be egoistic people take an instrumental orientation toward AI (as they tend to do for other humans; Guinote & Kim, 2020) but still hold the instrument accountable for failures or transgressions in their own self-enhancement and goal attainment. In other words, the AI fails as a tool for their own gain and responsibility needs to be delegated away from the egoist (i.e., offloading responsibility toward moral distancing; Hamman et al., 2010). In contrast, less egoistic individuals could see AI as more capable agents but also be more forgiving of them, given the primacy of equity and peace in their everyday decision-making—perhaps seeing the AI as flawed but not guilty in their joint interaction.

### 5.4 Limitations and Future Research

Our findings and interpretations are subject to the typical limitations inherent to the study design: We have subjected English-speaking Americans to a novel encounter with a researcher-constrained AI to complete an extrinsically motivated, light-weight task. Although we have mitigated these issues as justified in the method section, they should still be addressed through replication in other language and cultural contexts, repeated interactions with familiar AI, and across different tasks in naturalistic conditions. There are also other limitations of the current designs. We cannot verify that generating immoral content is *necessarily* seen as immoral behavior (i.e., a speech act) rather than badness inherent to the non-agentic words. As a test of priming language, this study accounts for only short-term, contextual language effects and do not account for how LAs could shape mental models over time (e.g., Higgins, 1996) or potentially influence the selection of information over time that could (dis)confirm or extend the prime (see Thibodeau & Boroditsky, 2011). Future research should parse and explore these potentials.

Additionally, we detected signals of other likely-important influences warranting exploration in future research. Much of our interpretation was data driven but still speculative and our propositions for egoism-instrumentalism links, proposed mechanisms around prime drift, horn effects, and mental-model influences require formal testing. Responsibility judgments were low overall and seem to be immune to priming or violations—given concerns over responsibility gaps in human-machine teaming, that phenomenon should be further explored. Finally, outside of the key findings reported and discussed, there are other significant patterns that emerge in data, however they are inconsistent and could be present or absent based on unknown sample differences; we also do not yet know how these process may translate to the messy moral landscape associated with AI-in-varied use, embedded in other contexts, and linked with specific message producers.

## 6 Conclusion

Across the four studies, we find one-shot priming human-AI interactions with lexical anthropomorphisms does not generally influence moral judgments of the AI, its behavior, or its responsibility when the AI behaves badly. Where there is influence, high-anthropomorphisms (those suggesting ontological rather than functional humanness) may be tapping into cognitively accessible notions of the AI as a potentially dishonest communicator. Prime drift (moving from the humanizing prime to a more tool-focused construal) suggests humans may engage established mental models for AI as a coping mechanism for behavioral expectancy violations. Observed immoral behavior, though, do offer exemplars for capacity perceptions—if an AI generates content about immorality (especially harm or contamination of human bodies), one is more likely to see it capable of immorality writ large. People centering egoistic values in everyday life may engage in moral distancing, simultaneously discounting an AI's ability to behave badly but still holding it responsible for bad behavior. Altogether, we interpret findings to indicate human judgments of novel AI bad behavior are less dependent on AI design or humanizing language, but instead negotiated *in situ* at the intersection of what we already know about AI, how the AI actually behaves, and how humans see themselves.

## ACKNOWLEDGMENTS

Research was sponsored by the Army Research Office and was accomplished under Grant Number W911NF-25-1-0079. The views and conclusions contained in this document are those of the authors and should not be interpreted as representing the official policies, either expressed or implied, of the Army Research Office or the U.S. Government. The U.S. Government is authorized to reproduce and distribute reprints for Government purposes notwithstanding any copyright notation herein. The authors thank Jon Stromer-Galley and Samiksha Singh for their development work on the stimulus AI.